# Applying Synchrotron Phase Measurement to the Estimation of Maximum Beam Intensity in the Fermilab Booster

Xi Yang and James MacLachlan

*Fermi National Accelerator Laboratory*

Box 500, Batavia IL 60510

**Abstract**

It is important to have experimental methods to estimate the maximum beam intensity for the Fermilab Booster as objective input into long term program commitments. An important existing limit is set by the available rf power. This limit is difficult to set *a priori*, because the real longitudinal impedance is not well known. The synchrotron phase at transition crossing was measured using both the mountain range plot and the direct phase measurement of the RF accelerating voltage relative to the beam, and results were consistent. They were applied to predict $6\times10^{12}$ maximum Booster beam intensity with present running conditions.

## 1  Introduction

A synchrotron phase detector has been implemented in the Fermilab Booster directly measuring the relative phase between the beam and the RF voltage in the RF cavities. It is a useful tool in numerous situations, such as in estimating the maximum beam intensity that can be achieved in Booster under the present running conditions, including the RF power, the efficiency of the LLRF system, etc.

## 2  Instrumentation

There are two RF input signals to the phase detector. One is the pickup from the resistive wall monitor in the straight section long 18 of the Booster ring. The signal to noise ratio from the beam current pickup is improved by connecting it through a bandpass filter because the 6 GHz bandwidth of the resistive wall monitor passes much other information besides the phase of the rf fundamental component. The other input signal to the phase detector is the vector sum of two downstream gap pickups, from station 7 of the group A and station 8 of the group B. It is necessary to have the summed input signal from both cavity groups because the two groups can be run with arbitrary but opposite offsets from the synchronous phase to control the effective rf amplitude acting on the beam. Cable lengths of the two input signals to the phase detector were carefully matched since Booster RF frequency changes from 37.8 MHz at injection to 53.4 MHz at extraction. Afterwards, the final adjustment of the two cable lengths was made using a criterion of zero degree synchrotron phase at injection. The phase detector delivers a signal of 9 degree per volt. A more complete technical description of the full phase detector system is being prepared (X. Yang and R. Padilla, work in progress). From the output of the phase detector, one can obtain the relative phase between the beam signal and the effective RF voltage at the gap of RF cavities.

## 3  Initial Application

The Booster beam intensity has been continually pushed to its limit in the last several years. A serious limiting factor has been the accelerating voltage available from the present RF system at the time of transition crossing. Due to the transition-crossing properties of a synchrotron, a bunch in Booster gets shorter as it approaches transition. It reaches its maximum peak current since the bunch length reaches its minimum near transition. The beam loses energy to the real part of the ring impedance, and the amount of the energy loss (deceleration) is closely related to the peak beam current. With greater beam intensity, the amount of the energy loss per particle at transition crossing due to the peak-current increment gets larger. The feedback system tries to correct this situation by driving the synchrotron phase toward 90 degrees to get more accelerating voltage. Until the synchrotron phase reaches 90 degrees the synchronous particle at least continues to receive the proper acceleration. However, the stable region of phase space around the synchronous particle (the bucket) shrinks as the synchronous phase increases. Any further increment of the beam intensity introduces an excessive reduction of the gap voltage, which can no longer be compensated by the change of the synchrotron phase. Beam is lost because of insufficient accelerating voltage. The process of the particle loss may continue until an equilibrium is reached. This can be understood as follows: during the fast ramp of a Booster cycle, the particles without correct acceleration will have different momentums with respect to the synchronous particle, such that their orbits are different from that of the synchronous particles. If the difference between the off-momentum particles and the synchronous particle is large enough to reach the dynamic aperture, these particles are lost. With more and more particles lost, the beam induced voltage will get smaller until an equilibrium is reached in which any remaining particles get the right acceleration. This problem is most severe at transition because the largest energy loss arises from the highest peak beam current in a Booster cycle, and average power demand is near its maximum because the ramp rate is also close to its maximum. A 90-degree synchrotron phase signifies that peak beam current has reached its limit.

## 4 Experimental Results

We used two different experimental approaches for the synchrotron phase measurement. One is to extract the synchrotron phase from a so-called mountain range plot (see Fig. 1) using the signal from the resistive wall monitor. The mountain range plot was used to record the process of the transition jump. The phase of the RF waveform changes from $\varphi$ to $180°-\varphi$ during the transition crossing. Thus the phase jump is calculated from Eq. 1.

$$\Delta\varphi = (180° - \varphi) - \varphi = 180° - 2\varphi \qquad (1)$$

The mountain range plot was triggered 80 µs before transition, with 1 trace per 8 turns for 40 traces. The time covered by the mountain range plot was more than 600 µs. We obtained the amount of the phase jump by measuring the jump ($\Delta t$) in time of the RF wave relative to one of the Booster bunches at the transition crossing and simultaneously recorded the period (T) of the RF wave. We can calculate the phase jump of the RF wave relative to the beam during the transition cross using Eq. 2.

$$\Delta\varphi = (\Delta t/T) \times 360° \qquad (2)$$

Substituting from Eq. 2 into Eq. 1, we obtain the synchrotron phase right before and after transition. Here, we assume the phase jump is symmetric about 90 degrees, meaning that the synchrotron phase changes from $\varphi$ to $180-\varphi$. However, this assumption is not exact in practice, as we will see later from the synchrotron phase measurement using the phase detector. From Fig. (1), the synchrotron phase near the transition approaches 90° more closely as the beam intensity goes higher. Assuming the per-particle energy loss per turn is linearly proportional to the intensity of the beam, we can estimate the maximum intensity that can be obtained in Booster when the RF power at the transition limits beam intensity. The result obtained from the data plotted in Fig. 2 is $6\times10^{12}$ protons per cycle.

The other approach to determining the synchronous phase is the direct measurement using the synchrotron phase detector. The results are shown in Fig. 3.

## 5 Conclusion

We can experimentally observe the relationship between the synchrotron phase and the beam intensity using the synchrotron phase detector. From this relationship we estimate the highest beam intensity that can be achieved in the Booster. Furthermore, we can use the synchrotron phase detector as one of the experimental tools to optimize the Booster running conditions by controlling the synchrotron phase after the transition crossing while keeping the beam intensity as high as possible. We can minimize beam loading and energy loss by reducing the peak current. One may try to make the feedback system faster and more efficient, etc. The effect of any such changes will be manifest in the synchrotron phase measurement.

**Acknowledgements**

Special thanks should be given to Chuck Ankenbrandt; he used his expertise and experience in suggesting the mountain range plot, which turned out to be quite successful. Bill Pellico supported this work by providing the phase shifted VCO as the trigger for the mountain range plot   Special thanks should be given also to Rene Padilla; he spent time together with authors on building the synchrotron phase detector. Milorad Popovic  provided helpful advice during the process of building the phase detector.

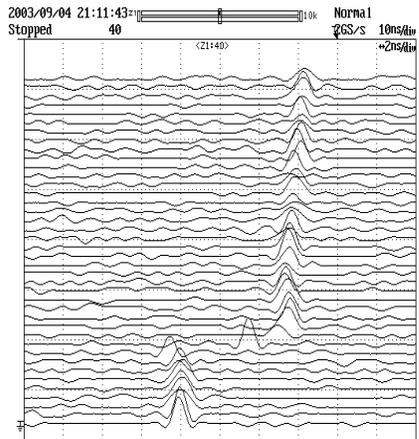
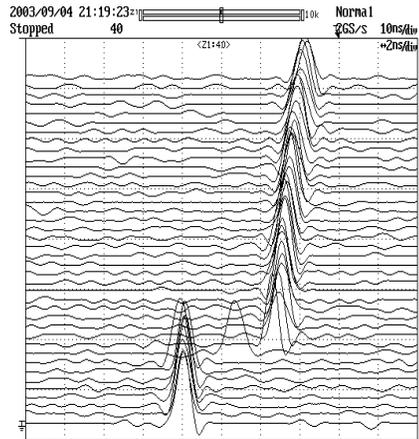
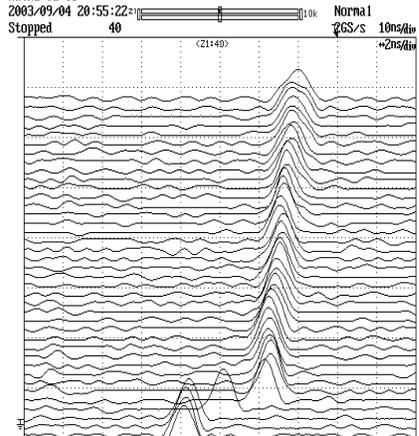
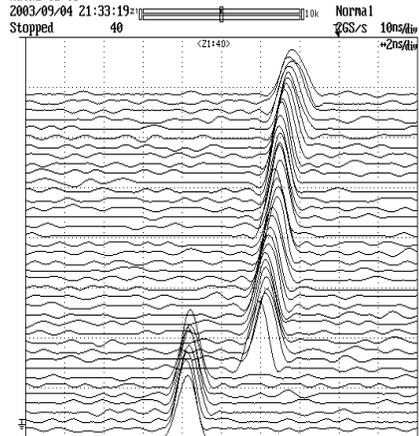
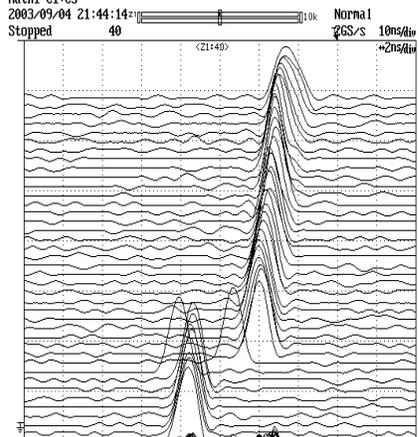
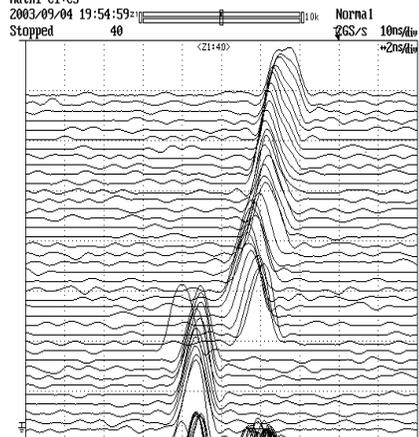

Fig. 1    5

Fig. 1 The mountain range plot was triggered 80 μs before the transition, and the same Booster bunch was plotted at one trace per 8 Booster turns. The same time scale (2 ns/div) for Fig. 1(a)-(f). During the plot time, the RF period is about 19.1 ns. The trigger for the mountain range plot was the LLRF driving signal of Booster RF cavities. In each plot, the time of the same booster bunch at 40 traces shows the relative position of the beam to the RF voltage. 1(a) the beam intensity is $0.4\times10^{12}$. 1(b) the beam intensity is $1.2\times10^{12}$. 1(c) the beam intensity is $1.9\times10^{12}$. 1(d) the beam intensity is $2.63\times10^{12}$. 1(e) the beam intensity is $3.5\times10^{12}$. 1(f) the beam intensity is $5.2\times10^{12}$.

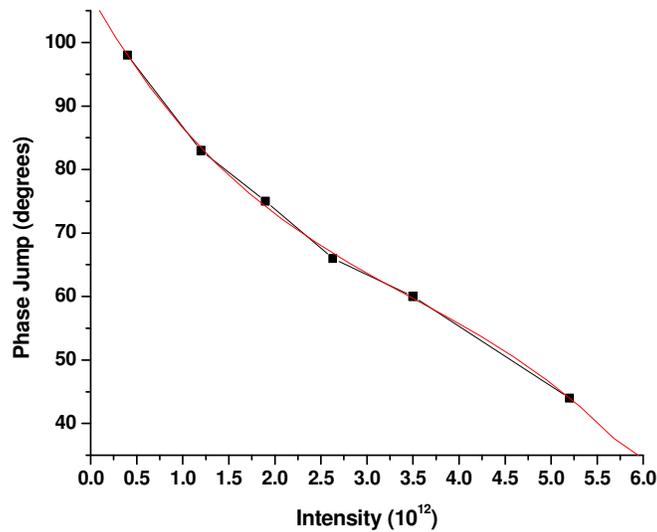

Fig. 2 Phase jump *vs.* the beam intensity. The red line is the polynomial fit of the experimental data to the 3$^{rd}$ order. From the fit, the maximum Booster intensity is predicted to be $6\times10^{12}$.

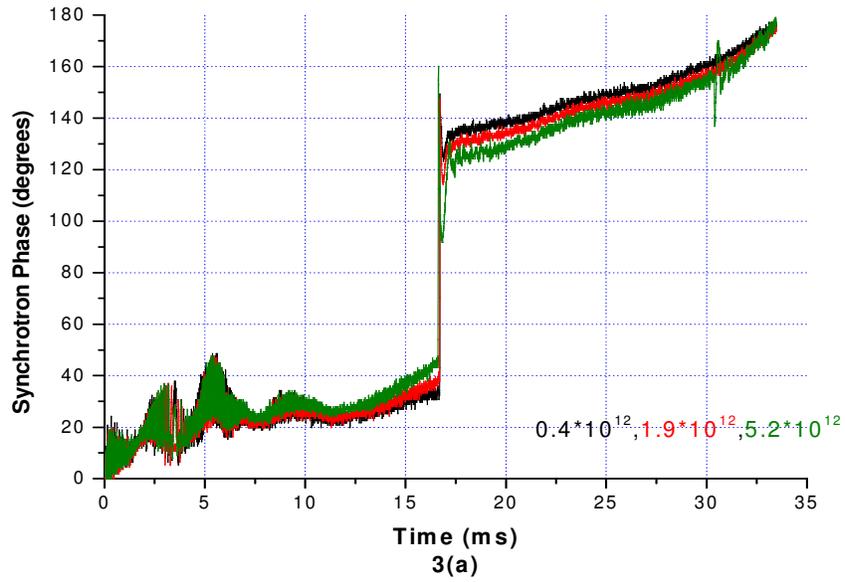

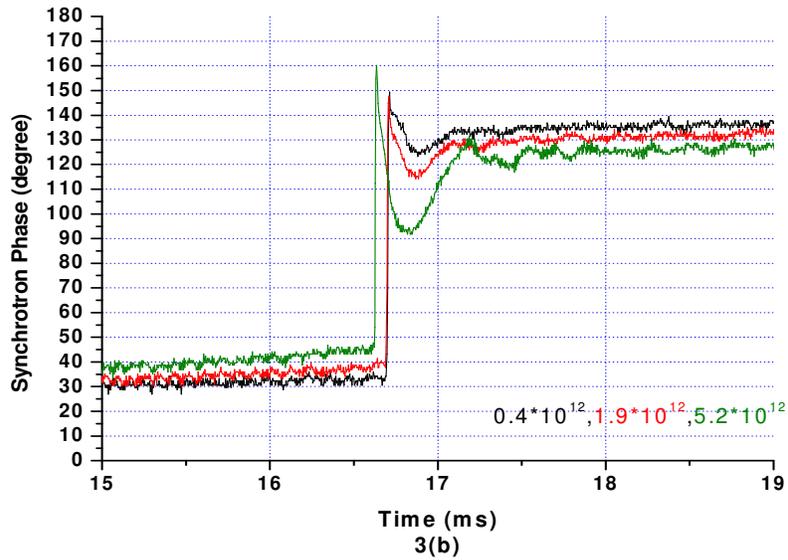

Fig. 3. The synchrotron phase measurement using the synchrotron phase detector. 3(a) the injection started at 0 ms. Three different beam intensities, $0.4\times10^{12}$, $1.9\times10^{12}$ and $5.2\times10^{12}$, were plotted. 3(b) the same as plot 3(a) with the expanded time scale at the transition crossing.